\documentclass{aa} %[referee]
\usepackage{graphics}      \usepackage[dvips]{graphicx}
\usepackage{array}         \usepackage{fancyhdr}
\usepackage{acromake}      \usepackage{subfigure}
\usepackage{amssymb}       %\usepackage{aas_macros}
\usepackage{natbib}
\bibpunct{(}{)}{;}{a}{}{,}

\begin{document}

% *** Standard A&A front page layout *******************************
\title{New aperture photometry of QSO 0957+561;\\
       application to time delay and microlensing}
\authorrunning{J.~E. Ovaldsen et al.}
\titlerunning{Photometry, time delay and microlensing for QSO 0957+561}

\author{J. E. Ovaldsen\inst{1} \and J. Teuber\inst{2} \and
   R. E. Schild\inst{3} \and R. Stabell\inst{1}}

\offprints{J.E.\ Ovaldsen, \email{j.e.ovaldsen@astro.uio.no}}

\institute{Institute of Theoretical Astrophysics, University of Oslo,
  P. O. Box 1029, Blindern, N-0315 Oslo, Norway 
  %j.e.ovaldsen@astro.uio.no, rstabell@astro.uio.no
  \and
  Centre for Advanced Signal Processing, Copenhagen, Denmark
  %jan.teuber@get2net.dk
  \and
  Harvard-Smithsonian Center for Astrophysics, 60 Garden Street,
  Cambridge, MA 02138, USA 
  %rschild@rudy.harvard.edu }
  }
\date{Received 29 September 2002 / Accepted 28 January 2003}

\abstract{We present a re-reduction of archival CCD frames of the
  doubly imaged quasar 0957+561 using a new photometry code. Aperture
  photometry with corrections for both cross contamination between the
  quasar images and galaxy contamination is performed on about 2650
  $R$-band images from a five year period (1992--1997).  From the
  brightness data a time delay of 424.9 $\pm$ 1.2 days is derived
  using two different statistical techniques. The amount of
  gravitational microlensing in the quasar light curves is briefly
  investigated, and we find unambiguous evidence of both long term and
  short term microlensing.  We also note the unusual circumstance
  regarding time delay estimates for this gravitational
  lens. Estimates by different observers from different data sets or
  even with the same data sets give lag estimates differing by
  typically 8 days, and error bars of only a day or two. This probably
  indicates several complexities where the result of each estimate
  depends upon the details of the calculation.
} % end Abstract
\maketitle

% *** Main text *****************************************************
\section{Introduction}
The first reported example of gravitational lensing, the twin quasar
QSO 0957+561, was discovered in 1979 by \citet{Walsh79}. It is one of
the most studied objects in modern cosmology, and the research and
monitoring campaigns have mainly been fueled by the desire to measure
the time delay, and thereby, to get an independent and direct estimate
of the Hubble parameter \citep{Refsdal64}. In addition, several groups
have tried to analyze the extrinsic variability in the light curves.
This variability is assumed to be caused by gravitational microlensing
(ML) by stars or MACHOs in the lensing galaxy (as predicted by
\citealp{Chang79}).

QSO 0957+561 is a doubly imaged quasar at a redshift $z=1.41$, with
the components A and B separated by about $6\farcs2$.\footnote{The
  literature consistently quotes $6\farcs1$, although accurate HST and
  VLBI astrometry yields values of $6\farcs169$ and $6\farcs175$,
  respectively \citep{Bern97}.} A massive, elliptical cD galaxy (named
G1) at $z=0.36$, located only $\simeq1''$ from the center of the B
image, seems to be the principal lensing object.

The closely juxtaposed quasar images and the extended brightness
profile of the lens galaxy make accurate photometry a challenge.
During the 1980s and mid-1990s, standard aperture photometry was
performed without any corrections for the light contamination between
the quasar components (crosstalk) or from the lens galaxy, see e.g.\ 
\citet{Schild86, Schild90, Kundic95}. Later reduction schemes have
tried to address the above-mentioned problems in order to reduce the
chance of correlated (seeing-dependent) brightness variations in the
light curves; e.g.\ \citet{CS99, CS00, Serra99}.  Precise photometry
is necessary for time delay determinations and for investigations of
possible microlens-induced fluctuations in the brightness records.

In spite of extensive observations by several groups, the time delay
($\tau$) between the two quasar images has proved hard to determine.
Complicating factors include heterogeneous data sets, large temporal
gaps in the data sets, and additional variability in one or both of
the quasar images (microlensing). Even more than 15 years after the
discovery, the time delay was not determined. However, there were two
favored candidates; $\sim~\!\!540$ days and $\sim\!\!~415$ days.  The
results of the different investigations prior to 1997 are summarized
in \citet{Haarsma97}.  \citet{Kundic97} convincingly settled the
long-standing controversy in favor of the lower value, finding $\tau =
417 \pm 3$ days. Since 1995 different groups have reported values of
$\tau$ in the range 416--425 days. The results seem again to
concentrate around two values; 417 days \citep{Kundic97, Pelt98, CS00}
and 424 days \citep{Pelt96, Oscoz97, Pijpers97, Serra99, Oscoz01}.

QSO 0957+561 was the first system to provide strong indications of
microlensing effects; uncorrelated brightness variations between the A
and B images were found by \citet{Vander89}. Several researchers have
reported microlens-induced variability in the quasar light curves.
\citet{Pelt98} found unambiguous evidence of long time scale (order of
several years) microlensing in the ``difference light curve'' (DLC; A
light curve minus time-shifted B curve).  Results are ambiguous when
it comes to the short time scale (lasting a few months) and rapid
(less than a few weeks) microlensing events.
\citet{Schild95}, \citet{Schild96} and \citet{CS00} have reported
interesting high-frequency features in the brightness record, having
amplitudes of only $0.03-0.05$ mag and time scales of months and even
weeks.  \citet{Goic98} also found fluctuations which could be
associated with microlensing events.  However, \citet{Schmidt98},
\citet{Wambs00} and \citet{Gil01} all found DLCs with no clear
microlensing signature, and notably no short time scale events with
$|\Delta m| > 0.05\,\textrm{mag}$ were observed. \citeauthor{Gil01}\ 
actually showed that the fluctuations in their DLC could be due to
(several) observational noise processes.  To reveal any rapid
fluctuations caused by microlensing, high quality images with good
temporal sampling are required.

This paper is mainly a summary of some results from a Master's thesis
project by \citet{Ovaldsen02} undertaken at the Institute of
Theoretical Astrophysics, University of Oslo, Norway.  We shall here
concentrate on the aperture photometry scheme, the time delay
estimation and microlensing investigation. The data set consists of
some 2650 archival CCD images of QSO 0957+561 covering a period of
nearly five years (June 1992 -- April 1997).  This data set has
previously been reduced by one of the authors (RES), but with cruder
corrections for crosstalk and galaxy contamination. In the next two
sections we discuss the data set and briefly present the main
principles of our photometry scheme. Then, from the final A and B
light curves (Sect.~\ref{S:phot_results}), the time delay is
determined using two different statistical techniques
(Sect.~\ref{S:time_delay}).  In Sect.~\ref{S:ML} we briefly
investigate the microlensing residual. The results are summarized and
discussed in Sect.~\ref{S:summary}. Our new photometry gives, among
other things, a time delay that differs significantly from the result
we obtain when employing the same method on the old RES brightness
data.

\section{Data set} \label{S:data_set}
RES and collaborators have monitored this lens system for over a
decade and amassed a large data set. Here we use a subset consisting
of around 2650 $R$-band CCD images taken with the 1.2m telescope at
Fred Lawrence Whipple Observatory atop Mt.\ Hopkins, Arizona, during a
five year period from June 1992 to April 1997. About 200 images were
discarded at an early stage due to various CCD defects, cosmic ray
hits, guiding errors, bad pre-processing etc. There are usually 4 or 6
frames per night.

Although taken with the same telescope, the quality of the frames
varies considerably. Cosmic rays, and especially bad pixels and bad
columns, occur frequently. Quite a few frames exhibit varying
background levels, not only in the form of a gradient across the
image, but as bright or dark ``patches'' at certain locations.  Such
frames may not have been properly calibrated. (Other artifacts from
poor pre-processing are also seen).

The image headers do not contain all the desired information, e.g.\ 
the pixel size and the gain factor are often missing. The gain is
fixed to 2.3 e$^{-}/$ADU. The pixel size is computed empirically for
each frame, using the calculated positions of typically 6 field stars
and the astrometry presented in the Guide Star Catalog II (GSC-II),
see Table~\ref{t:astrometry}.\footnote{The GSC-II is a joint project of
  the Space Telescope Science Institute and the Osservatorio
  Astronomico di Torino.}

Two different CCDs are employed. The scale of the first one is
approximately 0.65$''$/pixel (binned mode) and 0.32$''$/pixel
(unbinned), and that of the second one is 0.70$''$/pixel.  The range
of seeing values (FWHM) is approximately $1''-5''$, with a mean value
of around $2''$. The global background is mainly between 100 and 2000
ADU (92\% of the frames). The stellar images are typically
non-circular, the PSF having a mean ellipticity of 0.09, equivalent to
an axis ratio of 1.1.  We also note that the PSF often departs from
elliptical symmetry. The coma-like appearance is probably due to
tracking errors and astigmatism in the camera optics.

The sampling of the observations must be regarded as very good.
Besides the gaps in the summer months, the one day interval dominates.
More than 90\% of all time separations between consecutive observation
runs are less than eight days.

A very different and more homogeneous data set, comprising some 1000
$R$- and $V$-band frames obtained over four consecutive nights, is
discussed in a forthcoming paper \citep{Ovaldsen03}.

\section{Aperture photometry scheme} \label{S:phot}
The software used to reduce and analyze the CCD frames was developed
by JT and JEO. The entire package was written almost from scratch in
the Interactive Data Language (IDL)\footnote{A product of Research
  Systems, Inc.}; it is specially adopted to the 0957+561 twin quasar
system. Several sub-routines had been implemented by JT when working
on other quasar lens systems. Many of these remained unchanged.  All
steps are automated; from detection and localization of objects, via
field star photometry and calibration, to the actual quasar
photometry. We will only describe the main features of our photometry
scheme. The automatic source detection program, background
determination, centering algorithms etc.\ are not discussed here. We
refer to \citet{Ovaldsen02} for a complete and detailed treatment of
the entire package.

\subsection{Field star photometry and calibration} \label{S:star_phot}
\begin{figure}[!htb] %fig01
\sidecaption
\resizebox{\hsize}{!}{\includegraphics{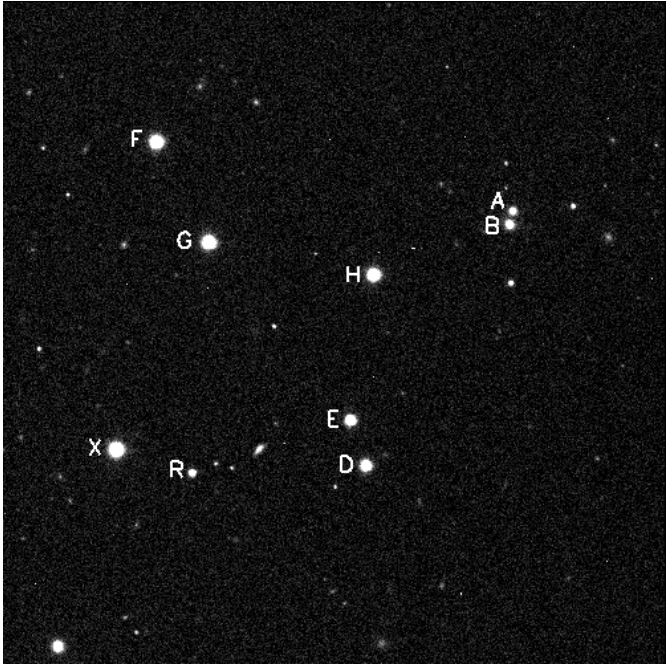}} % field_id.ps
\caption{The field surrounding QSO 0957+561A,B. North is up, east is
  left. The seven field stars (F, G, H, E, D, X and R) as well as the
  quasar components (A and B) are indicated.  The lens galaxy, not
  visible here, is located approximately $1''$ from B, slightly to the
  left of the vector from B to A. The frame is 4.7 arcmin on each
  side. (This particular image is taken with the 2.4m Hiltner
  telescope, Michigan-Dartmouth-MIT observatory, and is of better
  quality than the frames in our data set both in terms of seeing and
  spatial resolution; pixel size=0.275$''$/pixel, seeing $\approx
  1''$).}
\label{f:starfield}
\end{figure}

The separation between the two quasar images of $\sim 6''$ motivates
the use of $3''$ radius apertures. Using the same aperture size for
the comparison stars as for the target objects makes it easier to
transform the quasar intensities into standard magnitudes. We use the
stars F, G, H, E, D, X and R for the stellar photometry, see
Fig.~\ref{f:starfield} and Table~\ref{t:astrometry}.
% Figure
\begin{table}[!htb]
\caption{Reference star astrometry from GSC-II. Both right ascension,
  $\alpha$, and declination, $\delta$, are given in degrees. The
  rightmost column quotes the names of the stars as they appear in GSC-II.}
\begin{tabular}{cccl} \hline \hline
Star & $\alpha$ & $\delta$  & GSC-II id \\ \hline
F & 150.41117868 & 55.90692350  & N212232175   \\
G & 150.40026155 & 55.89505952  & N212232178   \\
H & 150.36551899 & 55.89125437  & N212232182   \\
E & 150.37040443 & 55.87405306  & N212232187   \\
D & 150.36712201 & 55.86871933  & N212232190   \\
X & 150.41965483 & 55.87058001  & N212232189   \\
R & 150.40374136 & 55.86788639  & N21223213199 \\ \hline 
\end{tabular}
\label{t:astrometry}
\end{table}
Of course, the use of such small apertures is only valid if they
collect the same fraction of the total light for \emph{all} point
sources, equivalent to the assumption that all point sources on a
frame have the same PSF.  We assume that this is approximately the
case. Preferably, one should observe reference stars with the same
spectral distribution as the primary targets; this would reduce the
error when calculating the magnitudes of the target objects. In the
case of 0957+561, the two quasar images are bluer than the field
stars. However, since we only have single band observations, we are
not able to correct for any color effects.

The large pixel size (mostly $\sim0.65''$/pixel) combined with the
relatively small apertures (radius=3$''$) give a quite irregular
polygon on the pixel array. To simulate a ``perfect'' circular
aperture we apply a weighting scheme for the pixels which lie on the
border. The value of a partial pixel is calculated as the original
pixel value multiplied by the ratio of the partial pixel area to the
total (square) pixel area. We also tried to quantify the implication
of a non-zero brightness gradient across the aperture border. This
second-order correction, however, proved insignificant.

The local background level is calculated from 20 small apertures
arranged in a circle of radius 20$''$ around the object of interest.
The apertures containing cosmic rays, bad columns, sources etc.\ are
automatically discarded.

We use seven comparison stars (Fig.~\ref{f:starfield}) to determine
the calibration level needed to put the quasar magnitudes on the
standard system. The instrumental intensities are compared to the
reference values, and any (5$\sigma$) outliers are registered. Some
frames contain fewer than seven comparison stars, but we require a
minimum of three to proceed. If there is more than \emph{one} outlier,
the frame is simply discarded. Our fundamental assumption is thus that
the intensities of \emph{all} the present stars (except one possible
outlier) should be consistent with the reference values. The
measurement errors are taken into account. With this procedure we make
sure that the calibration constant is calculated from ``well-behaved''
stars and, consequently, that the quasar magnitudes will be as
accurate as possible.

All measurement uncertainties (i.e.\ the standard deviation of the
aperture intensity $\hat{I}$) are calculated using the formula
\begin{equation} \label{eq:std_err}
\sigma (\hat{I}) = \sqrt{I/g + n_{a}\left(1+\frac{n_{a}}{n_{b}}\right)v_b} \ .
\end{equation}
$I$ is the background-subtracted source intensity (in ADU), $g$ is the
gain factor of the CCD, $v_b$ is the variance of the background, and
$n_a$ and $n_b$ is the number of pixels used in the determination of
the source intensity and background level, respectively. The $v_b$
parameter not only measures the variance of the sky level, but also
fluctuations (inhomogeneities) due to faint background sources and the
CCD readout noise. Hence, $v_b$ is always larger than the Poisson
variance of the sky level.

\subsection{Photometry of the two quasar components} \label{S:DQ_scheme}
As previously mentioned, we use 3$''$ radius apertures centered on
each quasar image. These apertures will be subject to seeing- (and
ellipticity-) dependent light contamination from the neighboring
quasar component and the underlying G1 lens galaxy.

\subsubsection{Galaxy subtraction}
To correct for the galaxy's light contamination, we decided to
subtract from each frame a synthetic model of G1. Upon request, G.\ 
Bernstein kindly provided surface brightness data obtained from HST
observations in the $V$-band and Kitt Peak observations in the
$R$-band (see \citealp{Bern97}).  In order to find the color offset
for the HST data, we simply looked for what gave the best agreement in
the overlap area of the profiles. The ``correction'' $V-R=1.3$ seemed
to merge the profiles well. Although the study by \citet{Bern97}
indicated an ellipticity gradient and isophote twist, we decided to
model the galaxy with fixed values for the ellipticity (0.28) and
position angle (53 degrees) -- compare with their Fig.~2.

The position of the center and the orientation of the semi-major axis
are calculated from the positions of the quasar images and from the
relative astrometry of \citet{Bern97}. To synthesize the galaxy we
start by oversampling the pixels four times.  The value of each
sub-pixel is computed by interpolating the brightness profile
quadratically. The ellipticity and position angle are taken into
account. When determining the calibration (or zero) level for the
galaxy, the small $3''$ apertures do not suffice. G1 is an extended
object whose profile is much broader and totally different from that
of the stars.  For this reason it is important to calibrate a resolved
object like G1 with the ``total'' light from the comparison point
sources, here taken to be the flux in apertures of radius $12''$.

The model is finally ``smeared out'' in accordance with the seeing on
each particular image. This is done by convolving the synthesized
galaxy with the image PSF. Having performed the proper scaling and
positioning, the convolved galaxy image is simply subtracted from the
frame.

\subsubsection{Crosstalk correction}
Several methods to minimize cross contamination between the A and B
images were explored, some of which were similar to the procedure in
\citet{CS99}. However, because we have to calculate the PSF for each
frame (used in the modeling of the synthetic lens galaxy), we decided
to utilize one of the characterizing features of the PSF-fitting
technique. The A and B images are cleaned from the frame in an
iterative fashion, thereby allowing aperture photometry to be
performed on each quasar image after the galaxy is subtracted and
\emph{after} the neighboring twin is cleaned from the frame. The
cleaning works well for a wide range of seeing conditions, and this
way of eliminating the crosstalk between the A and B images proved to
be significantly more robust than the other methods (it is, for
instance, less sensitive to bad columns, bad pixels etc.).

\section{Photometric results} \label{S:phot_results}
\subsection{Stellar photometry}
\begin{figure}[!htb] %fig02
\sidecaption
\resizebox{\hsize}{!}{\includegraphics{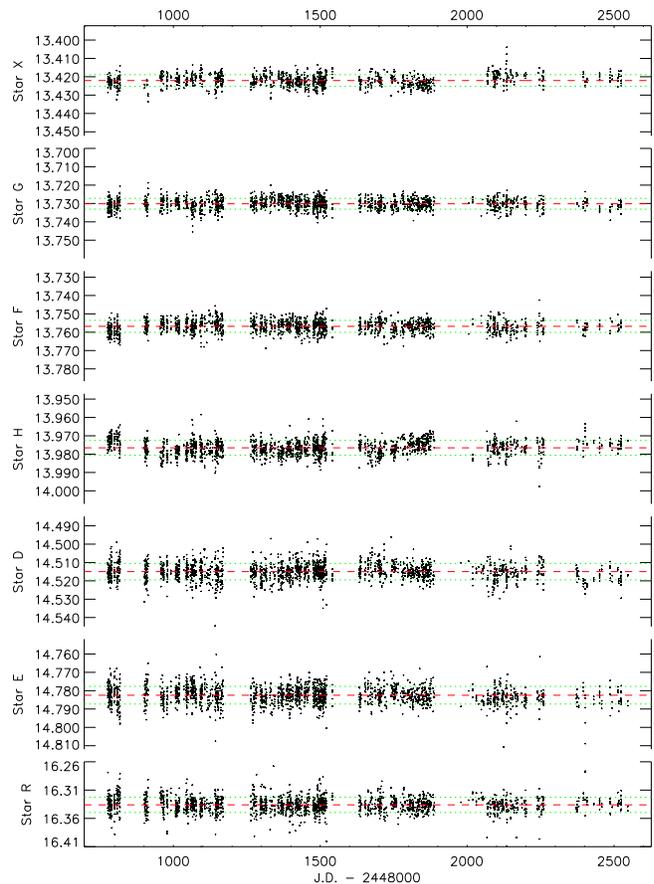}} % Starmags_ap.ps
\caption{Aperture photometry of the field stars around QSO 0957+561.
  Plotted are the $R$-band magnitudes of the reference stars from all
  accepted data frames, as a function of Julian Day. Dashed red line
  is the mean value, dotted green lines are the $1\sigma$ limit. The
  stars are arranged in order of decreasing brightness. Note that the
  scaling for the R star is different from the others.}
\label{f:star_phot}
\end{figure}

2486 frames were analyzed with respect to reference star photometry,
and the calibration procedure (see Sect.~\ref{S:star_phot}) accepted
2028 frames.  Fig.~\ref{f:star_phot} shows the magnitudes of the seven
reference stars for all \emph{accepted} frames, as a function of
Julian Day (J.D.). Table~\ref{t:star_stat} shows some statistics for
each star.
% Figure
\begin{table}[!htb]
\caption{Statistics of the field star aperture photometry. For each
  star the mean magnitude and standard deviation are tabulated, along
  with the mean of the individual formal errors, and reference
  magnitude.}
\begin{tabular}{@{}ccccc@{}} \hline \hline
Star & Mean $R$-mag & $\sigma_\mathrm{mag} $ & Mean formal error & 
Ref.\ mag \\ \hline
 F  &  13.757  &  0.0033  &  0.0030  &  13.757 \\
 G  &  13.730  &  0.0029  &  0.0029  &  13.730 \\
 H  &  13.977  &  0.0040  &  0.0032  &  13.977 \\
 E  &  14.783  &  0.0048  &  0.0047  &  14.782 \\
 D  &  14.515  &  0.0045  &  0.0041  &  14.515 \\
 X  &  13.422  &  0.0032  &  0.0027  &  13.422 \\
 R  &  16.336  &  0.0135  &  0.0123  &  16.335 \\ \hline
\end{tabular}
\label{t:star_stat}
\end{table}
We remark that the stars F, G, H, E, D, X and R were saturated on 280,
289, 97, 0, 2, 719 and 0 frames, respectively.

From each frame we only calculated the magnitudes of the comparison
stars which were unsaturated and had intensities consistent with the
reference values. The calibration constant was computed from the same
stars. As we see from the plots, there are a few ``outliers''.  Some
of these can be identified as points with large error bars, but some
are true outliers in the sense that they should have been discarded.
The calibration constant is not necessarily biased by such outliers in
all cases, because it is derived from several (3--7) stars.

As expected, the scatter in magnitudes increases for fainter stars.
The R star is by far the faintest, and consequently has the largest
standard deviation for the calculated magnitudes, i.e.\ 13.5 mmag.
Given that its brightness is $\sim0.3$~mag greater than the quasar
images, this should indicate the \emph{minimum} general dispersion to
be expected in the quasar light curves due to photometric noise. After
all, the photometry of the two quasar images is much more complicated
than that of an isolated star; the galaxy subtraction and the cross
talk correction obviously increase the error budget of images A and B.

\subsection{Quasar photometry}
\subsubsection{Binning and censoring}
The number of data points (or accepted frames) per night, $n$, varies
from one to six. The observation were made within a small time
interval, thus it seems appropriate to combine all data points for a
particular night, and only quote ``binned'' magnitude values. It is
almost impossible to perform a rigorous statistical analysis with so
few data points (and an unknown number of outliers due to erratic
photometry of the twin images). We decided to address this issue in a
simple and transparent way. From the image A magnitudes and the image
B magnitudes on each night, we computed the corresponding
\emph{median} values. With this approach, single outliers do not bias
the results too much, at least for $n \geq 3$.  Obviously, for $n$ = 1
or 2, the median-filtering will not throw away possible outliers.
However, without any a priori knowledge, this is about the best we can
do. We quote error bars which correspond to the median points.

Before the median filter was applied we censored the data, accepting
only images with seeing $\leq 3''$ FWHM, background level $\leq$ 3000
ADU, PSF axis ratio $\leq$ 1.3 and AB-separation $6\farcs175 \pm
0\farcs05$. The final data set was then reduced to 1720 images. The
subsequent ``binning'' yielded 422 data points for each quasar image.

\subsubsection{Crosstalk and galaxy contamination}
Aperture photometry on images A and B was performed both with and
without the corrections for crosstalk and galaxy contamination (see
Sect.~\ref{S:DQ_scheme}), so that we could check the performance. We
now make a few comments on the results from analyzing the 1720
accepted images.

\begin{figure}[!htb] %fig03
\sidecaption
\resizebox{\hsize}{!}{\includegraphics{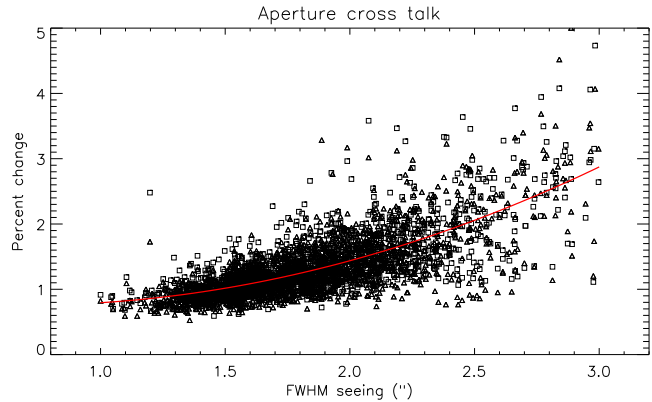}} % crosstalk.ps
\caption{The percent change in the A and B aperture flux when correcting
  for crosstalk, as a function of seeing. The crosstalk from the B
  (A) image into the A (B) aperture is marked with open squares
  (triangles). The corrections are virtually the same for the two
  apertures. A best fit quadratic curve is overplotted.}
\label{f:crosstalk}
\end{figure}

Fig.~\ref{f:crosstalk} displays how the light from the two quasar
images affects each aperture.
% Figure
We emphasize that this effect is not only a function of seeing; the
ellipticity is certainly also a factor. In particular, consider an
image where the PSF is highly elliptical and has a semi-major axis
parallel to the line joining the centers of A and B. The crosstalk
would be larger here compared to the case where the semi-major axis is
perpendicular to the AB vector. (In fact, the scatter of the
corrections can be significantly reduced by imposing stricter limits
on the ellipticity).

For completeness, a least squares second order polynomial fit was
computed using all the data points (see also \citealp{CS00}). The
formula for the intensity corrections, $\delta I$, reads
\begin{equation} \label{eq:CT_formula}
\delta I(s) = \left[0.944 - 0.546s + 0.397s^2\right]\% \ ,
\end{equation}
where $s$ = FWHM ($\arcsec$).
As can be seen in the figure, the curve fits the data quite well.  The
scatter is partly due to ellipticity, but some of the deviant points
are a result of bad cleaning of the quasars images; on some frames the
flux in the measuring aperture is still affected by the neighboring
quasar image, which has not been properly removed/subtracted.

\begin{figure}[!htb] %fig04
\sidecaption
\resizebox{\hsize}{!}{\includegraphics{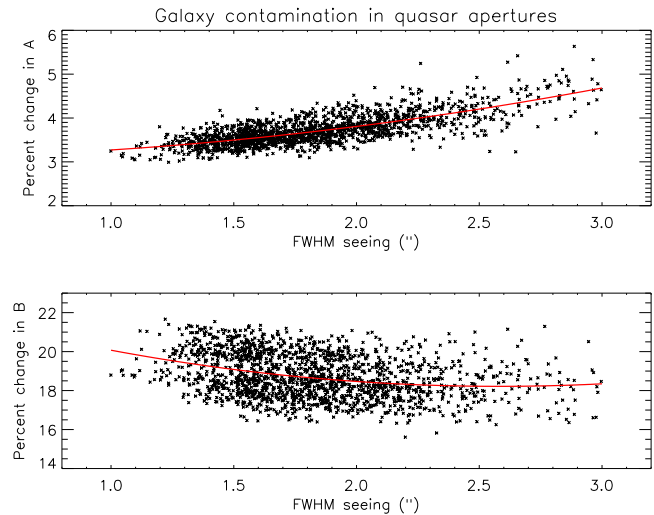}} % gal_cont.ps
\caption{Light contamination (in percent) in the A and B apertures due to
   the G1 lens galaxy, as a function of seeing. Best fit quadratic
   curves are also plotted.}
\label{f:gal_cont}
\end{figure}

We also measured the flux in the quasar apertures before and after the
galaxy model had been subtracted. Fig.~\ref{f:gal_cont} shows the
light contribution from the galaxy to the A and B apertures as a
function of seeing. Note that crosstalk between A and B has already
been ``eliminated''.
% Figure
As seeing deteriorates, galaxy light systematically seeps out of the B
aperture, but into the A aperture.  Best fit quadratic curves are
overplotted to guide the eye.  Some of the scatter in the plots is due
to the fact that A and B itself varied during the observational period
(the G1 contribution is compared to the A and B fluxes on each
particular frame.)  The contribution to the A image aperture is
between 3 and 5\%, and has a moderate scatter. The B image aperture
has corrections of roughly $20-18\%$ due to the G1 galaxy. Here, the
scatter is rather large, having a ``full amplitude'' of $\pm$2\%.  It
does not seem to increase with seeing.  This probably indicates that
the calibration/zero level for the galaxy is determined equally well
(or poor!) for the whole range of seeing conditions. Comparable
corrections for subtracting the lens galaxy contribution were
discussed in detail by \citet{CS00}, whereas the original RES
reductions incorporated subtraction of a fixed 18.34 magnitude
correction for the lens galaxy contribution to the B aperture flux.

\subsubsection{Quasar light curves}
\begin{figure*}[!htb] %fig05
\centering
\includegraphics[width=17cm]{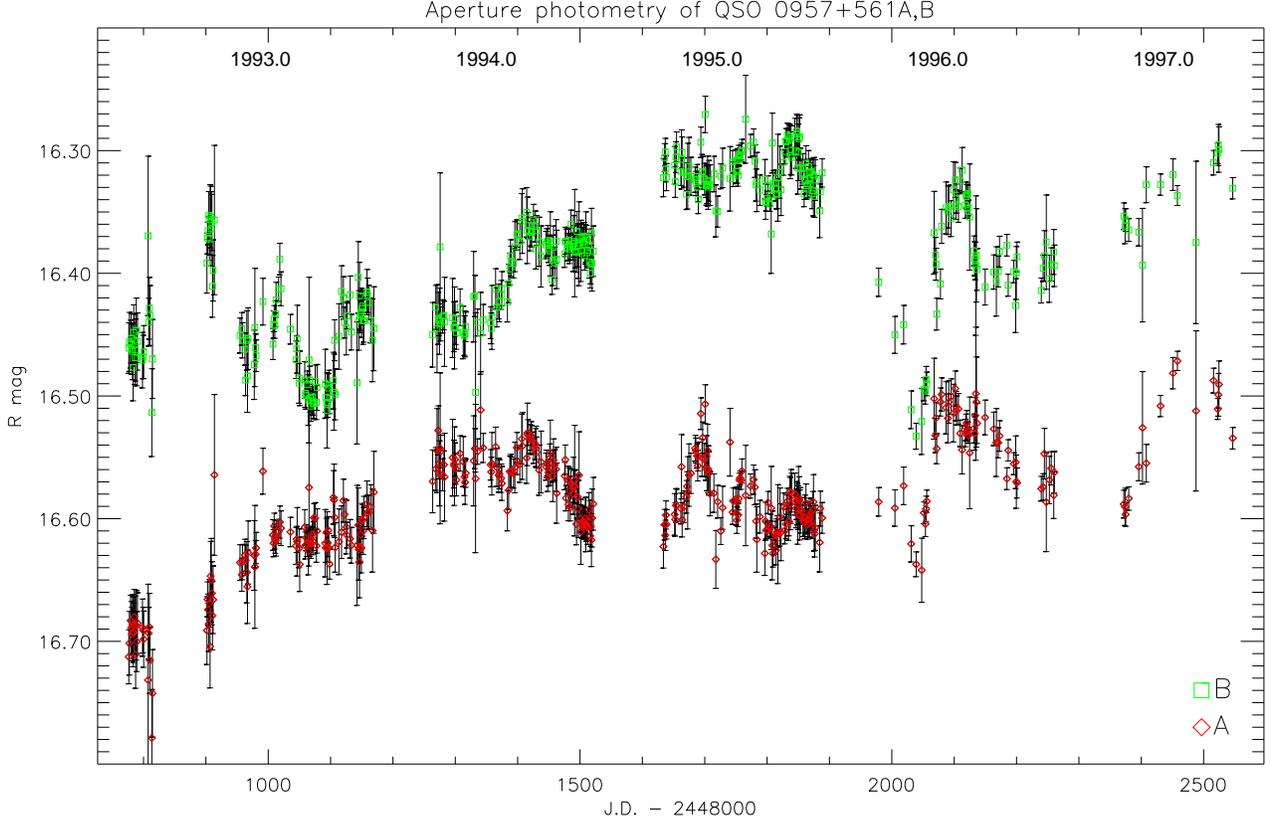} % ABmag_ap.ps
\caption{Results from aperture photometry of QSO 0957+561A,B. There
are 422 binned data points for each quasar image. The B data is
shifted by $-0.15$ mag. Error bars are $1\sigma$ limits.}
\label{f:ABmags_ap}
\end{figure*}

Fig.~\ref{f:ABmags_ap} displays the light curves of QSO 0957+561A,B
corresponding to the period June 1992 to April 1997.\footnote{The data table is
available at \texttt{http://www.astro.uio.no/$\sim$jeovalds/DQmags.html}.}
% Figure
We note that the light curves show variability on both short (order of
weeks) and long (order of years) time scales.  For some periods there
is also an apparent \emph{``zero lag'' correlation} between A and B.
This is best seen for J.D.--2448000 $\gtrsim$ 1700.  Such a
correlation should in general not exist, because the signal from image
B lags A by some $\tau\sim 420$ days. We have not been able to
identify the cause of this frame-by-frame correlation. It has also
been reported by \citet{Colley03}, and is always presumed to be some
systematic effect caused by errors in the photometry; however the
amplitude exceeds any known error source.

Although our ``binning'' scheme only uses the nightly median magnitude
value for each quasar image, we can still estimate the dispersion for
nights with two or more accepted frames. The mean standard deviations
of the magnitudes on each night are 12 mmag and 11 mmag for A and B,
respectively. We note that the mean of the \emph{formal} error bars,
as seen in Fig.~\ref{f:ABmags_ap}, is 17 mmag for both quasar images.
The formal error bars are rather conservative, as they include the
formal errors from Poisson statistics, galaxy subtraction and
calibration (see \citealp{Ovaldsen02} for details).

We also made a rough and simple estimate of the day-to-day dispersion
within the A and B brightness data: First each light curve was
smoothed with a 7-point median filter (making sure not to filter over
gaps greater than three days). Then the original data (A or B) was
subtracted from the corresponding median-filtered curve. We allowed a
maximum time gap of 1.5 days between two data points to be subtracted.
The residuals should thus probe fluctuations in the A and B light
curves on this time scale. We assume that this very short time scale
variability is not dominated by microlensing effects. The standard
deviations of the residuals are $\sigma^\mathrm{resid}_\mathrm{A}$ =
11 mmag, and $\sigma^\mathrm{resid}_\mathrm{B}$ = 10 mmag. These
values are quite consistent with similar estimates for the image set
made by \citet{Schild95B}, who found 9.5 and 12.0 mmag; however their
reductions lack the corrections for aperture crosstalk and galaxy
subtraction that are strictly functions of seeing, and are more
susceptible to systematic errors on time scales relevant to seeing
changes.

\section{Time delay} \label{S:time_delay}
The main analysis is performed using a method based on dispersion
estimates, but we also explore a different technique based on $\chi^2$
minimization. We use the data corresponding to Fig.~\ref{f:ABmags_ap}.

\subsection{Dispersion estimates} \label{S:disp_est}
The algorithm for the Dispersion estimation technique is included in
the ISDA (Irregularly Spaced Data Analysis) package, designed by J.\ 
Pelt to perform various tasks on irregularly spaced time series.  It
is discussed extensively in \citet{Pelt94,Pelt96}, so we provide here
only a short review.  The principle of the Dispersion method is simply
to measure the dispersion, $D^2$, between the A and B image light
curves for different time shift values, $\tau$.  The true value should
show up as a \emph{minimum} in the dispersion spectrum, $D^2(\tau)$.
By dispersion we here mean the sum of the squared differences between
the A and the B image magnitudes (see below).

The data model is
\begin{eqnarray*}
A_i & = & q(t_i) + \epsilon_A(t_i)\ , \ i=1,\ldots,N_A \\
B_j & = & q(t_j-\tau) + l(t_j) + \epsilon_B(t_j)\ , \ j=1,\ldots,N_B \ ,
\end{eqnarray*}
where $q(t)$ denotes the inherent quasar variability, which should be
the same in the two images. $l(t)$ takes care of the unknown
amplification ratio between A and B, as well as any long time scale
microlensing. The observational errors are $\epsilon_A(t_i)$ and
$\epsilon_B(t_j)$.

The combined light curve (denoted in the formulae as $C$) is constructed
by taking the $A$ values as they are and ``correcting'' the $B$ data
by $l(t)$ and shifting them by $\tau$: 
\begin{equation}
C_k(t_k) = \left\{
  \begin{array}{ll}
    A_i        & \mathrm{if}\ t_k=t_i  \\
    B_j-l(t_j) & \mathrm{if}\ t_k=t_j+\tau \ ,
  \end{array}
\right.
\end{equation}
where $k=1,\ldots,N$ and $N=N_A+N_B$. The dispersion of the combined
light curve (abbreviated as CLC in the text) is now computed for
a range of $\tau$-values. The resulting dispersion spectra
\begin{equation} \label{eq:disp_spc}
D^2(\tau) = \min_{l(t)} D^2(\tau,l(t))
\end{equation}
can subsequently be inspected with regard to minima. The time shifts
of the most significant minima are candidates for the true time delay.

The accuracy of the observations is taken into account by using the
statistical weights, $W_i=1/\epsilon_A(t_i)$ and $W_j=1/\epsilon_B(t_j)$.
The squared difference between two data points in the CLC (see
estimates below) must be multiplied with the combined statistical
weights $W_k = W_{i,j} = \frac{W_i W_j}{W_i+W_j}$.

With $l(t)$=constant, the A and B curves are considered to be
unaffected by microlensing variability, differing only by the unknown
ratio of the amplification factors of gravitational (macro-) lensing.
We shall, however, also compute spectra where we account for slowly
varying microlensing effects in one or both of the light curves. In
these cases we set $l(t)$ equal to a polynomial (typically of order
two to eight). The B data is thus ``modified'' by the perturbing
polynomial, into $B_j + l(t_j)$, and the coefficients of the
polynomial are determined in such a way as to minimize the dispersion
between A and B data.

In this analysis we use two different methods to estimate
dispersions. The simplest one is
\begin{equation} \label{eq:disp3}
D^2_3 = \min_{l(t)}\frac{\sum_{k=1}^{N-1} S_k W_k G_k (C_{k+1}-C_k)^2}
                        {2\sum_{k=1}^{N-1} S_k W_k G_k} \ ,
\end{equation}
where $W_k$ is the statistical weights, and $G_k=1$ only if $C_{k+1}$
and $C_k$ are from different data sets. (That is, one point from A and
one from B. Otherwise, $G_k=0$). $S_k$ constrains the time gap between
the AB or BA pairs;
\begin{equation}
S_k = \left\{
  \begin{array}{ll}
    1 & \mathrm{if}\ |t_{k+1}-t_k| \le \delta  \\
    0 & \mathrm{if}\ |t_{k+1}-t_k|  >  \delta \ ,
  \end{array}
\right.
\end{equation}
where $\delta$ (measured in days) is the largest time separation
allowed, also called the \emph{decorrelation} length.

The second statistic is
\begin{equation} \label{eq:disp4g}
D^2_{4,h} = \min_{l(t)}\frac{\sum_{n=1}^{N-1}\sum_{m=n+1}^{N}
  S_{n,m}^{(h)} W_{n,m} G_{n,m}(C_n-C_m)^2}
  {\sum_{n=1}^{N-1}\sum_{m=n+1}^{N} S_{n,m}^{(h)}W_{n,m} G_{n,m}} ,
\end{equation}
where $W_{n,m}$ are statistical weights and $G_{n,m}$ controls that
only AB or BA pairs are considered, just as in the previous estimate.
In this scheme more pairs are included by not only considering
strictly neighboring pairs; $S_{n,m}^{(h)}$ weights $(C_n-C_m)^2$
according to the corresponding temporal separation, $|t_n-t_m|$. We
may use a flat window ($h=1$), linear ($h=2$) or Lorentzian ($h=3$)
down-weighting, see \citet{Pelt96}. Here we use linear
down-weighting;
\begin{equation} \label{eq:wscheme2}
S_{n,m}^{(2)} = \left\{
  \begin{array}{ll}
    1-\frac{|t_n-t_m|}{\delta} & \mathrm{if}\ |t_n-t_m| \le \delta  \\
    0 & \mathrm{if}\ |t_n-t_m|  >  \delta \ .
  \end{array}
\right.
\end{equation}

The $D^2_3$ and $D^2_{4,2}$ estimates calculate the dispersion %only
between A and B points, with an upper limit on the corresponding time
separation.  $D^2_3$ includes only consecutive AB (or BA) points, and
does not involve any smoothing. The $D^2_{4,2}$ estimate has the
advantage of including much more pairs, and thus suppressing noise in
the dispersion spectra. However, one should be careful not to
over-smooth the spectra by using large decorrelation lengths. We shall
employ the $D^2_{4,2}$ estimate most of the time.  Different values of
the decorrelation length, $\delta$, will be tested, as well as
various models (constant versus higher-order polynomials) to account
for the additional (microlens-induced) variability.

ISDA contains a simple bootstrap procedure for calculating the error bars
for the minima in the dispersion spectra. The CLC is smoothed and the
corresponding residuals (data points minus ``smoothed curve values'')
are re-shuffled 1000 times to create bootstrap samples. Smoothing is
performed using a 7-point median filter, with an upper limit
(typically a few days) on the time separation between two successive
data points.

The time delay value from a particular dispersion estimate is taken to
be the mean of the time delay distribution (an example is given in
Fig.~\ref{f:td_hist}).  The standard deviation of the distribution
gives the estimated error. We quote the 1$\sigma$ errors.

\subsubsection{Complete data set}
From the complete data set of 422 data points for each of the two
quasar images, we discarded six outliers.

\begin{figure}[!htb] %fig06
\resizebox{\hsize}{!}{\includegraphics{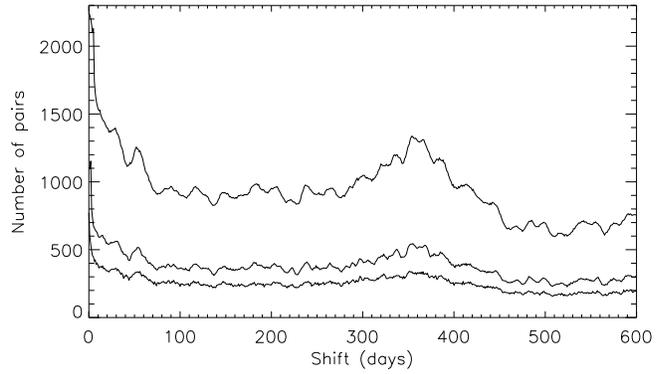}} % wind_funct.ps
\caption{Window functions for the complete data set when employing the
  estimates $D^2_{4,2}$ ($\delta=5$ days; upper curve, $\delta=2$
  days; middle curve) and $D^2_3$, $\delta=5$ days (lower curve).}
\label{f:win_ap5}
\end{figure}

The number of pairs included in the dispersion estimates depends on
$\delta$. Fig.~\ref{f:win_ap5} displays selected window functions for
$D^2_3$ and $D^2_{4,2}$. The window function is the number of nearby
AB (or BA) pairs in the CLC as a function of time shift.
% Figure
Obviously, larger $\delta$ yields more pairs in the computation, but
the overall shape of the window functions remains more or less the
same (the curves get smoother as $\delta$ increases).  The sampling of
the observations may disfavor some time shifts, i.e.\ the number of
pairs of nearby points in the CLC can be very low for certain shifts.
Fortunately, there are no major depressions in the curves, so the
statistical reliability of the dispersion values should not vary much
for the different trial shifts (especially in the interesting range
400--440 days). This is a reassuring and important fact.

\begin{figure}[!htb] %fig07
\resizebox{\hsize}{!}{\includegraphics{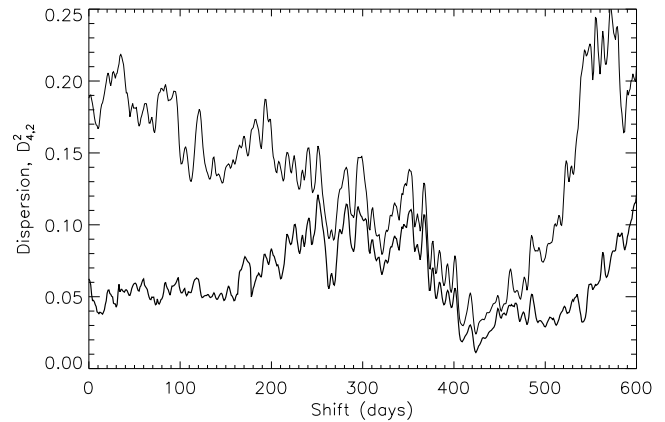}} % Ap_full_d3_BP.ps
\caption{$D^2_{4,2}$ dispersion spectra, $\delta=3$. The thin curve
  corresponds to $l(t)=l_0$, while the thick curve is computed by also
  accounting for additional fluctuations ($l(t)$ is a 5th order polynomial).}
\label{f:ap_full}
\end{figure}

We shall plot the dispersion spectra for trial shifts in the range
380--480 days. However, before we present and discuss the behavior of
the dispersion curves in this limited range, we show in
Fig.~\ref{f:ap_full} a plot of two spectra calculated using the
$D^2_{4,2}$ estimate where the interval goes from 0--600 days.
% Figure
Over the entire range, the dispersion is smaller for the estimate
that includes the perturbing polynomial. A higher-order polynomial
would account even more for differences in the two quasar signals, and
thus decrease the general dispersion even further. One must be wary
not to ``over-correct'' the B data, though.

\begin{figure}[!htb] %fig08
\resizebox{\hsize}{!}{\includegraphics{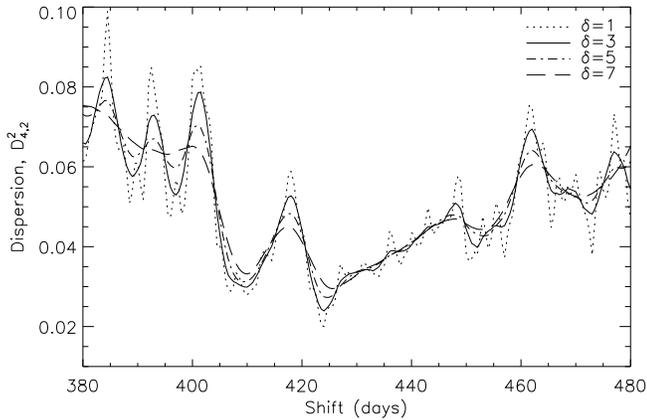}} % AP_d1357_wB.ps
\caption{$D^2_{4,2}$ dispersion spectra with $\delta$ = 1, 3, 5, 7 days. No
  corrections for additional, possibly microlens-induced variability.}
\label{f:ap_d1_11_B}
\end{figure}

\begin{figure}[!htb] %fig09
\resizebox{\hsize}{!}{\includegraphics{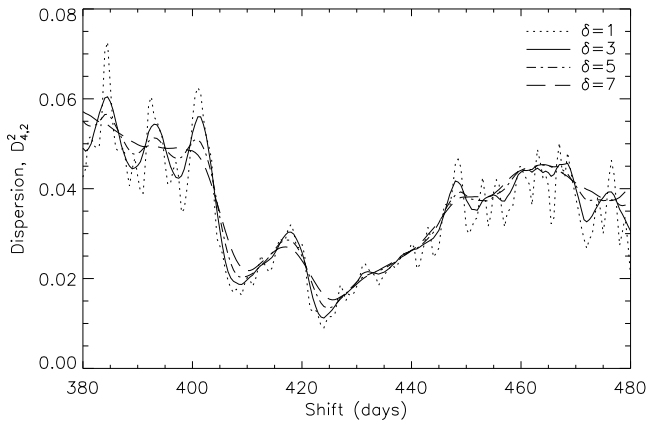}} % AP_d1357_p6.ps
\caption{$D^2_{4,2}$ dispersion spectra with $\delta$ = 1, 3, 5, 7
  days. A 5th order polynomial was used for modeling of additional
  variability in the light curves.}
\label{f:ap_d1_11_p6}
\end{figure}

We first computed dispersion spectra using various decorrelation
lengths, but without any corrections for microlensing ($l(t)=l_0$).
Then, the calculations were repeated, but this time we included
polynomials to model long time scale ML variability in the light
curves. The results were not very sensitive to the degree (2nd--8th
order) of the perturbing polynomial. To limit the number of variables,
we fixed the degree to 5th order.  Figs. \ref{f:ap_d1_11_B} and
\ref{f:ap_d1_11_p6} display a selection of spectra ($\delta$ = 1, 3,
5, 7 days) derived from the two methods.
% Figure
% Figure

The curves are smoother for larger $\delta$. Note that $\delta=1$
hardly involves any smoothing. For a given $\delta$, the two methods
(without/with correction for additional variability) yield very
similar results. In both cases, the position of the minimum increases
slightly for increasing decorrelation length, from 424 to 425.5 days.
Secondary minima are almost always found around 410 days, but they are
moderated as the smoothing increases.

\begin{figure}[!htb] %fig10
\resizebox{\hsize}{!}{\includegraphics{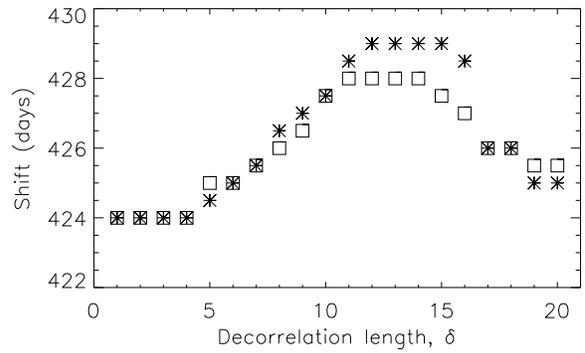}} % Ap_delta_spc.ps
\caption{Dependence of the $\delta$ parameter for the $D^2_{4,2}$
  statistic. Asterisks: no correction for additional variability. Open
  squares: 5th order polynomial.}
\label{f:delta_spc_ap}
\end{figure}

We also computed spectra for $\delta=1-20$ with increments of one, and
noted the position of the corresponding minima.
Fig.~\ref{f:delta_spc_ap} plots shift values corresponding to minimum
dispersion as a function of the $\delta$ parameter.
% Figure
Here we see more clearly that the minima are shifted towards higher
values as pairs with larger time separations are included in the
estimates.  However, for $\delta > 14-15$, the trend is reversed, but
we are probably smoothing too much already.
Because the sampling of the observations is rather good (85\% of the
time separations are less than five days), we can afford to use small
decorrelation lengths.  This is reassuring, because it reduces the
danger of bias from pairs with large time separations.
A striking feature is also seen: whether we account for slowly varying
microlensing effects or not, the minima are all at 424 days for
$\delta$ = 1, 2, 3 and 4.

The same procedure was employed for the $D^2_3$ statistic, which only
includes \emph{consecutive} A and B points whose time separation in
the CLC is less than $\delta$. Hence, no smoothing is performed. These
spectra were similar to the curves in Figs.~\ref{f:ap_d1_11_B} and
\ref{f:ap_d1_11_p6} corresponding to $\delta = 1$. The results
confirmed the trends and features which were highlighted above.  Since
the number of pairs included in this estimate is lower than the
``smoothing'' estimates, it is not as reliable, statistically
speaking. The spectra did exhibit more noise, but consistently
produced global minima between 423.5 and 424.5 days when $\delta$ was
$\leq 4$ days. As before, larger decorrelation lengths yielded higher
time shift values.

To summarize: the different dispersion estimates ($D^2_3$,
$D^2_{4,2}$, both with and without microlensing correction) all
produce minima which are concentrated around 424 days for $\delta \leq
4$. It seems that larger decorrelation lengths introduce bias. Hence,
we performed bootstrap runs only for the limited range of
$\delta$-values.  As noted earlier, the results were not significantly
affected by the nature of the polynomial used to model additional
variability. In particular, for the preferred range of $\delta$ (i.e.\ 
1--4 days), the minima in the dispersion spectra were all at 424 days
for degrees of order two to eight. We do not want to use very
high-order polynomials, as this could suppress some of the intrinsic
quasar fluctuations. Hence, it seems justified to fix the degree to
five.

\begin{figure}[!htb] %fig11
\resizebox{\hsize}{!}{\includegraphics{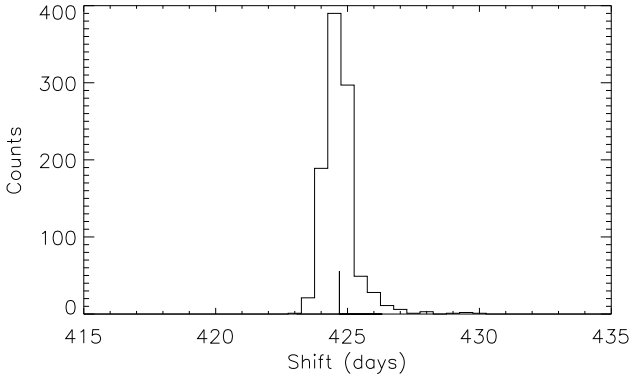}} % hist_ap_d4_p6.ps
\caption{Bootstrap results. The mean time delay from 1000 bootstrap
  runs is indicated by the large tick mark. Statistic: $D^2_{4,2}$,
  $\delta=4$, $l(t)$ is a 5th order polynomial.}
\label{f:td_hist}
\end{figure}

Table~\ref{t:td_results_ap} lists the results of the bootstrap
procedure.
\begin{table}[!htb]
\caption{Time delay results for the $D^2_{4,2}$ dispersion estimate applied
  to the complete aperture photometry data set. A 5th order polynomial
  was used to model additional variability in the quasar light curves.
  As noted in the text, other polynomials gave very similar results.
  Estimated errors are 1$\sigma$ limits.}
\begin{tabular}{ccc} \hline \hline
Statistic    & $\delta$ & Time delay (days) \\ \hline %\hline
$D^2_{4,2}$  &  1       & 424.8 $\pm$ 1.6  \\
$D^2_{4,2}$  &  2       & 424.3 $\pm$ 0.6  \\
$D^2_{4,2}$  &  3       & 424.8 $\pm$ 1.3  \\
$D^2_{4,2}$  &  4       & 424.7 $\pm$ 0.7  \\ \hline
\end{tabular}
\label{t:td_results_ap}
\end{table}
In Fig.~\ref{f:td_hist} we show an example of the distribution of
time delays from one of the bootstrap runs.
% Figure
Here, the mode and the median were both 424.5 days, while the mean
time delay was 424.7 days. The shapes of the other distributions were
similar, and they all had a small skew. We also note that the $D^2_3$
estimate gave similar results with the bootstrap procedure.  The
estimated time delay using different setups of the Dispersion
estimation technique agree well. We thus take the most probable time
delay to be the average of the numbers in the table, i.e.\ 424.7 days,
with a mean estimated error of 1.1 days. However, we do not claim that
the true error is as low as this.

The magnitude difference, $\Delta m_\mathrm{AB}$, between the A and B
components in the 1992--1997 time span was $\approx 0.076$ mag (we
were not able to compute error bars for this parameter). Hence, B was
somewhat brighter than A in the time span which our data covers. This
is commonly explained by microlensing in the B component, see e.g.\
\citet{Pelt98}.

Finally, it is worth noting that \emph{all} spectra show a local
\emph{maximum} at $\sim$418 days. So with this particular data set and
the Dispersion estimation method, we can say that a time delay of
roughly 418 ($\pm$ 2) days seems rather unlikely.

\subsubsection{Truncated data sets} \label{S:trunc_data}
We now estimate the time delay using only selected data segments.  We
divide each light curve into four parts, corresponding to the seasons
with J.D.$-$2448000 roughly in the ranges 850--1200 (period~1),
1200--1600 (period~2), 1600--1900 (period~3) and 1900--2300
(period~4), see Fig.~\ref{f:ABmags_ap}. Assuming a time delay of
around 425 days, it is clear that there is (fortunately) a large
overlap between periods 1, 2 and 3 of A and periods 2, 3 and 4 of B,
respectively. We thus have the possibility of estimating the time
delay between the quasar images from three different data subsets (let
us call these $S1$, $S2$, $S3$). The motivation behind this is to see
whether the truncated data sets all produce a minimum in the
dispersion spectra around 425 days. We shall not perform an exhaustive
analysis, though.

Because the number of pairs included in the calculations is much lower
for these subsets, the statistical reliability is not as good as in
the case where we used the complete data set. We employ the
$D^2_{4,2}$ estimate, $\delta$ in the range 2--5 days, and compute
spectra for trial shifts in the range 400--450 days. The effect of
correcting for any non-intrinsic quasar fluctuations is also tested.
For this we use (only) a 3rd order polynomial (the curves overlap for
about 200 days in all three cases, and we do not allow for any
extrinsic high-frequency components within this time span).

\begin{figure*}[!htb] %fig12
\sidecaption
\includegraphics[width=12cm]{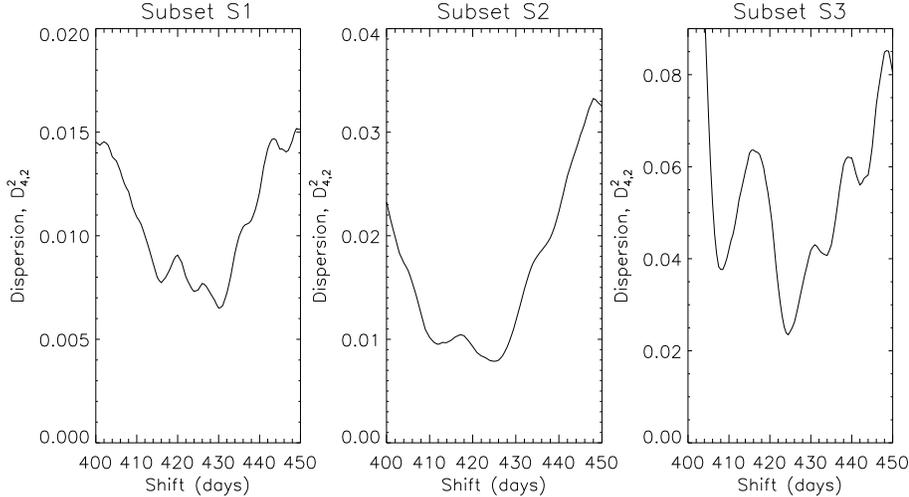} % Ap_seasons_d4_wB.ps
\caption{$D^2_{4,2}$ dispersion spectra. $\delta = 4$. No corrections
  for microlens-induced variability. The shapes of the dispersion
  spectra from the different data sets are very different. The position of
  some minima correspond to minima in the window function, i.e.\ the
  number of pairs in the dispersion estimates reaches a minimum for
  these time shift values.}
\label{f:Ap_seasons}
\end{figure*}

Fig.~\ref{f:Ap_seasons} displays $D^2_{4,2}$ dispersion spectra
computed for the subsets $S1$, $S2$ and $S3$, assuming a constant
magnification ratio, $l(t)=l_0$.
% Figure
Allowing for a time-dependent magnification ratio (to account for
possible microlensing-effects) did not significantly change the
overall shape of the curves. The minima are sometimes split in two
for short decorrelation lengths, so we use $\delta=4$ which reduce
the ``noise''.

For the first data subset, $S1$, the deepest minimum is at 430 days.
Secondary minima are seen around 424 and 415 days. We checked the
window function too, and it contained a prominent minimum for the 430
day shift. It might be that the 430 day candidate is caused by the
combination of irregular sampling and an ``unfortunate'' time shift.
The second subset, $S2$, yields two minima in the dispersion spectrum,
412 days and 425 days, the latter being the deepest. Here, the window
function had no unfavorable time shift. The third plot shows the
results using the $S3$ data, and here the deepest minimum is
positioned around 425 days.  Secondary minima are found for time
shifts of 408 and 434 days. The window function had minima at 408 and
425 days, corresponding exactly to two of the observed local minima in
the dispersion spectrum.

Bootstrap runs were performed to get an estimate of the uncertainty.
From the distribution of 1000 time delays, the results (mean and
standard deviation) were as follows: 430.6 $\pm$ 2.7 days ($S1$), 424.9
$\pm$ 3.0 days ($S2$), 426.1 $\pm$ 2.3 ($S3$). We do not attempt to
judge the reliability of the different time delay candidates. It is
interesting to see, however, that the local minima may be found in
usually one of three regions, i.e.\ around 410, 425 and 430 days. Also, a
local peak in the interval 416--420 days is found in all spectra, thus
supporting the statement made in the previous section that the correct
time delay is less likely to lie in this range.

For the complete data set, we found in the previous section that
$\Delta m_\mathrm{AB} \approx 0.076$ mag. However, a brief look at the
combined light curve (B shifted in time by $-$425 days and in
magnitudes by 0.076 mag) indicated that a single magnitude difference
did \emph{not} optimally align the A and B data. We thus computed
$\Delta m_\mathrm{AB}$ for each of the three data subsets in order to
investigate this further. We got $\approx0.089$ mag, $\approx0.086$
mag and $\approx0.050$ mag for subsets $S1$, $S2$ and $S3$,
respectively. The fact that the magnitude difference seems to be
time-dependent will be discussed in Sect.~\ref{S:ML}.

\subsubsection{Time delay from RES's data}
Now that we have performed a time delay determination using our new
photometric results, it might be interesting to see whether the
``old'' reductions by Schild and collaborators\footnote{See data table
  at \texttt{http://cfa-www.harvard.edu/$\sim$rschild/}.} give similar
results. We used 537 data points for each quasar image which covered
the \emph{same} period as our data. An extensive analysis is beyond
the scope of this paper. We shall thus only
comment on the main results from the $D^2_{4,2}$ dispersion estimate.

\begin{figure}[!htb] %fig13
\resizebox{\hsize}{!}{\includegraphics{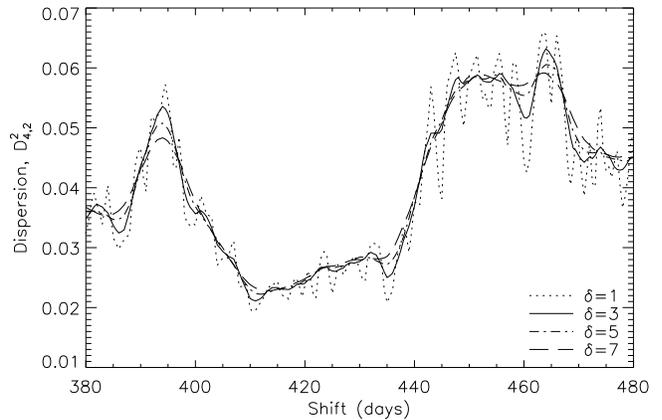}} % Rudy_d1357_p6.ps
\caption{Dispersion spectra $D^2_{4,2}$ ($\delta$~= 1, 3, 5, 7 days) with
  5th order polynomial, using RES's data.}
\label{f:Rudy_spc}
\end{figure}

The results revealed a few interesting general trends:
\begin{itemize}
\item With short decorrelation lengths ($\delta \leq 5$ days) and no
  correction for microlensing, the dispersion spectra had minima at
  435 days.  Prominent, but marginally higher minima were seen at
  $\sim$412 days.
  By introducing polynomials, $l(t)$, of varying degrees to model
  (hypothetical) microlensing, the global minimum was shifted to
  around 411 days for the same range of $\delta$.
\item For larger decorrelation lengths, the minima were found in the
  region 411--415 days, \emph{irrespective} of the exact nature of
  $l(t)$ (constant versus higher-order polynomials). Increasing
  $\delta$ from 6 to 12 days consistently produced minima at larger
  time shifts, going from 412 up to 415 days.
\end{itemize}
We present in Fig.~\ref{f:Rudy_spc} dispersion spectra $D^2_{4,2}$
with $\delta$~= 1, 3, 5 and 7 days.
% Figure

The second point above describes a general trend seen in all the
different analyzes : the position of the dispersion minimum increases
with increasing decorrelation length.

Here we find that the position of the global minimum was 411 days for
$\delta \leq 4$ days, and it did \emph{not} depend on the order of the
perturbing polynomial (2nd to 8th order). There are certainly
indications of additional variability in the observational data, hence
it seems justified to introduce the $l(t)$ polynomial into the
dispersion minimization process. The results of the bootstrap
procedure are listed in Table~\ref{t:td_Rudy}. The mean time delay is
411.7 days, with a mean estimated error of 1.9 days.
\begin{table}[!htb]
\caption{Time delay results for the $D^2_{4,2}$ dispersion estimate applied
  to the RES photometry data. A 5th order polynomial was used to
  model additional variability in the quasar light curves. Estimated
  errors are 1$\sigma$ limits.}
\begin{tabular}{ccc} \hline \hline
Statistic    & $\delta$ & Time delay (days) \\ \hline
$D^2_{4,2}$  &  1       &  411.4 $\pm$ 2.3  \\
$D^2_{4,2}$  &  2       &  411.8 $\pm$ 1.8  \\
$D^2_{4,2}$  &  3       &  411.8 $\pm$ 1.8  \\
$D^2_{4,2}$  &  4       &  411.8 $\pm$ 1.7  \\ \hline
\end{tabular}
\label{t:td_Rudy}
\end{table}

\subsection{$\chi^2$ minimization} \label{S:chi2_min}
The method of \citet{Burud01} is based on $\chi^2$ minimization
  between the data and a numerical model light curve. Microlensing in
one or both light curves may be corrected for. We shall in the
following carry out a brief, non-exhaustive time delay analysis with
this method. Because the procedure is explained in detail in the paper
by \citeauthor{Burud01}, we only summarize the main features.

The underlying idea is that, in the absence of ML, one can model the
two quasar light curves with \emph{one} model curve, $g(t)$, together
with two parameters, $\tau$ and $\Delta m$, describing the time shift
and magnitude offset between images A and B. An arbitrary model curve
with equally spaced sampling points is $\chi^2$ minimized to the two
original light curves. The minimization is done only for the observed
data points, so only the model curve is interpolated and not the data.
Because the data is irregularly sampled and contain noise, a smoothing
scheme is necessary. Here, the model light curve, $g(t)$, is smoothed
on a time scale, $T_1$, corresponding to the typical sampling interval
of the data.  The smoothing term is multiplied by a Lagrange
parameter, $\lambda$, which can be chosen so that the model curve
matches the data correctly in a statistical sense for adopted Gaussian
statistics (this parameter has no physical meaning). In addition, each
data point is given a weight which depends on the relative distances
to all other points in the curve.  Down-weighting is performed using a
Gaussian with FWHM = $2\sqrt{2 \ln 2}\,T_2$, where the user may choose
the $T_2$ parameter.  The weights are normalized, so that the maximum
value of $W_i$ is 1.  This would be the case if only one point is
within the time interval defined by $T_2$. According to the authors, a
sensible choice of $T_2$ would be the approximate time scale of the
intrinsic quasar fluctuations.

The method was only applied to the complete light curves. We did not
attempt to model higher-order ML fluctuations in the light curves, as
this is quite an elaborate process. A wide range of parameter setups
was tested, and the results proved to be remarkably stable. We present
only the main results.

\begin{figure}[!htb] %fig14
\resizebox{\hsize}{!}{\includegraphics{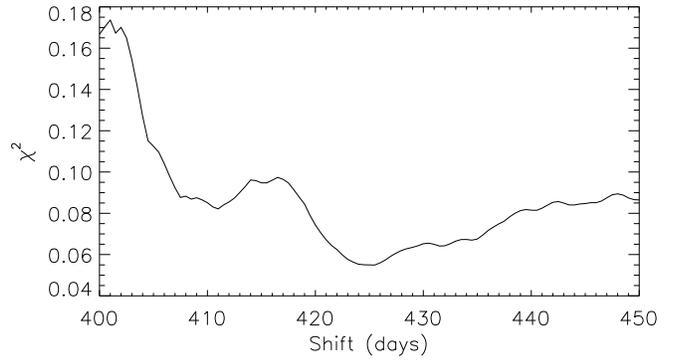}} % Ap_g4.ps
\caption{$\chi^2$ as a function of time shift.}
\label{f:chi_ap}
\end{figure}

The $\chi^2$-values as a function of time shift (in the range 400--450
days) are plotted in Fig.~\ref{f:chi_ap}. The parameters were as
follows: $T_1=4$ days, $T_2=20$ days and $\lambda=4000$.
% Figure
We recognize some of the features from the analysis with the
Dispersion estimation technique: The minimum $\chi^2$-value occurs for
a time shift of 425 days.  A secondary minimum is found around 411
days, but the $\chi^2$ is not as low as the tiny, local minimum at
431--432 days. The overall shape of the $\chi^2$ distribution remains
the same even for large variations in the parameters. The lowest
$\chi^2$-value is always in the range 424--426 days.

The results form Monte Carlo simulations yielded a time delay of
425.1 $\pm$ 1.3 days. Also with this method we find that a time delay
of roughly 415--420 days seems less likely; the $\chi^2$-curve
typically has a maximum in this range.

\section{Microlensing} \label{S:ML}
We will now briefly investigate the microlensing residual in the
quasar light curves. The standard procedure is to shift the light
curves in time to correct for the different light travel times, and
then subtract them from each other. The last step is not trivial, as
the A and B data points are irregularly sampled.  This means that when
we shift the B data in time by $-\tau$, the A and B points will
generally not overlap. The B data point to be subtracted from a
particular A point might be several days away. We have addressed this
issue in a simple way.

After having shifted the B curve by $-$425 days, we check for each
data point of image A whether there is a point from the B curve within
a certain \emph{gap limit} of the current A point. If this is the
case, then B is subtracted from A, and the result is stored in a
``residuals array'' with an averaged time argument.  The procedure
only makes use of the original, raw data points, and there is one free
parameter, namely the gap limit. The exact value of this parameter
depends mostly on the spacing of the data, but also on the (assumed)
ML time scale. With good sampling the gap limit can be set quite low
(a few days) and thus, at least in principle, enable investigation of
rapid ML.  Lowering the gap limit will obviously decrease the number
of points in the residuals array. On the other hand, the $A-B$
differences are then calculated from AB pairs with smaller time
separations, which is a good thing if one wants to probe short time
scale fluctuations in the residuals.

With a time delay of $\sim$425 days, our A and B data overlap for
about 3.5 years. The ML investigation will consequently only cover
this time span. The large temporal gaps in the residuals data (a
result of the lack of observations in the months where 0957+561
was below the horizon) also precludes a continuous ``signal'' for the
whole period.

\subsection{Trends on long time scales} \label{S:l}
\begin{figure}[!htp] %fig15
\resizebox{\hsize}{!}{\includegraphics{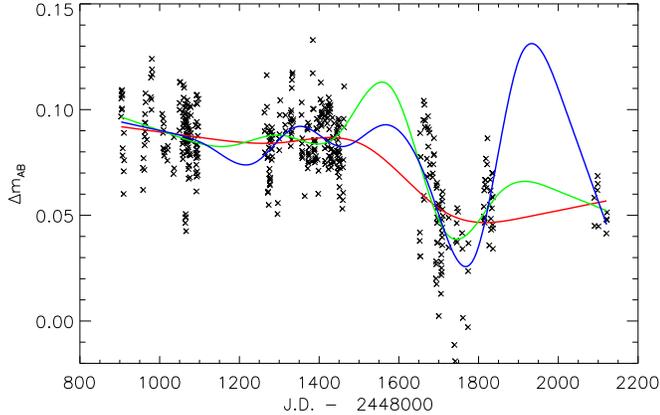}} % ML25_splines.ps
\caption{Time-shifted B curve values subtracted from A curve
values. Dates on the abscissa relate to the (unshifted) A light
curve. The $A-B$ differences, $\Delta m_\mathrm{AB}$, are clearly
time-dependent. Three examples of spline approximations to the ML
residuals are also shown. Red, green and blue lines correspond to 6,
9, and 11 nodes, respectively.}
\label{f:AB_residuals}
\end{figure}

In Fig.~\ref{f:AB_residuals} we display the $A-B$ residuals ($\Delta
m_\mathrm{AB}$), computed by following the above procedure. We adopted
a time delay of 425 days, and the gap limit was set to 2.5 days.
% Figure
There are three seasons which contain an adequate number of points.
The results look rather noisy, but there are some significant
features. In particular, the third season clearly has a different
magnification ratio between A and B, compared to the first two.
Moreover, it also varies within the particular season.  Variability on
shorter time scales may also be seen at certain periods.

The amplitude of the variations in the first two seasons is about 0.05
mag. From the plot we can see that the average magnitude offset for
the two first seasons are $\sim$0.09 mag, while for the third period the
number is roughly 0.05 mag. This is in good agreement with the values we
obtained in Sect.~\ref{S:trunc_data}. % for the truncated data sets.

Here, we only want to get an idea of the general trends in the ML
residuals. We thus tried fitting standard cubic splines with different
number of nodes into the residuals, see Fig.~\ref{f:AB_residuals}.
The (red) curve with 6 nodes ``detects'' only changes on long time
scales. It is thus a rather conservative guess as to how microlenses
in the macro-lensing galaxy affected the light from the quasar images.
The general trends are similar to the results of \citet{Pelt98} --
compare with the three last ``seasons'' in their Fig.\ 9.
\citeauthor{Pelt98}\ do not have the points around J.D.--2448000=2100
which we do (see Fig.~\ref{f:AB_residuals}). Although very sparse,
these data indicate that the curve does not continue to fall off.  The
other two splines probe finer details in the ML residuals, and are
more optimistic approximations. Some of the variability, notably in
the gaps, is highly questionable.  Still, the $\sim$0.05 mag drop in
the difference data from the second to the third season (on the order
of 300 days) seems significant. We conclude that microlensing
variability of approximately 5\% amplitude on time scales of less than
a year has been significantly observed.

\subsection{Short time scale variation} \label{S:s}
\begin{figure}[!htb] %fig16-18
\resizebox{\hsize}{!}{\includegraphics{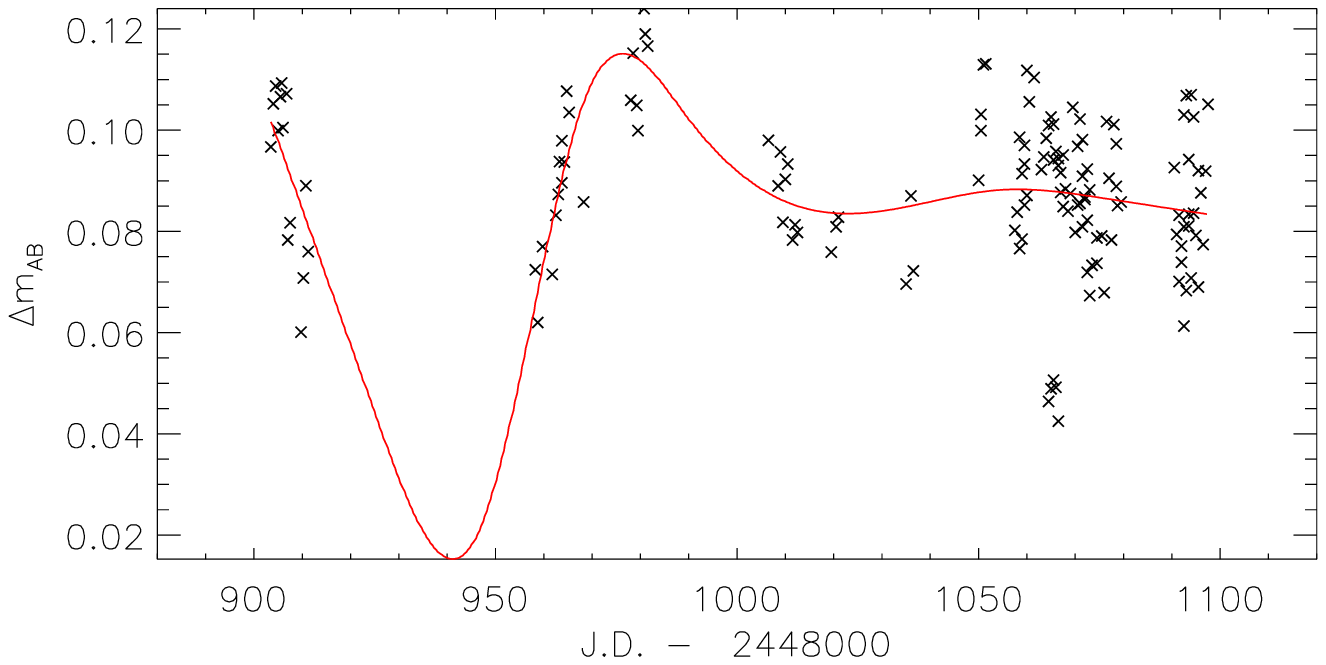}} % regr_fit25_1.ps
\resizebox{\hsize}{!}{\includegraphics{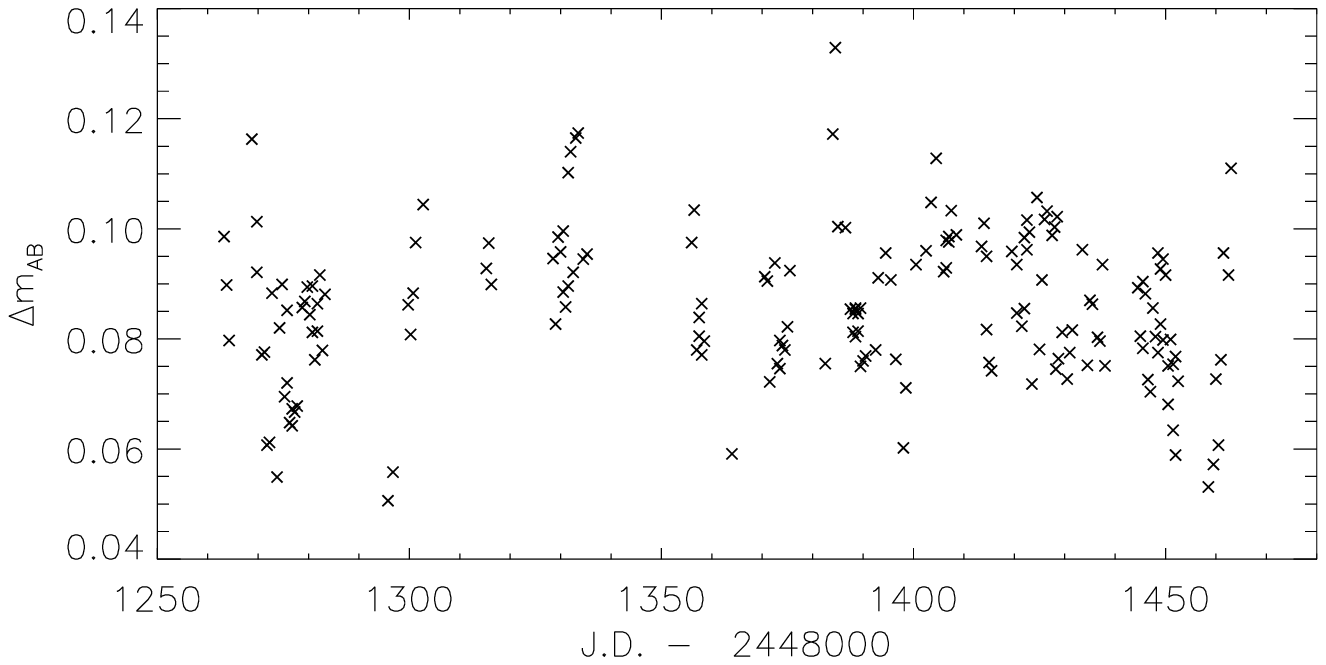}} % regr25_2.ps
\resizebox{\hsize}{!}{\includegraphics{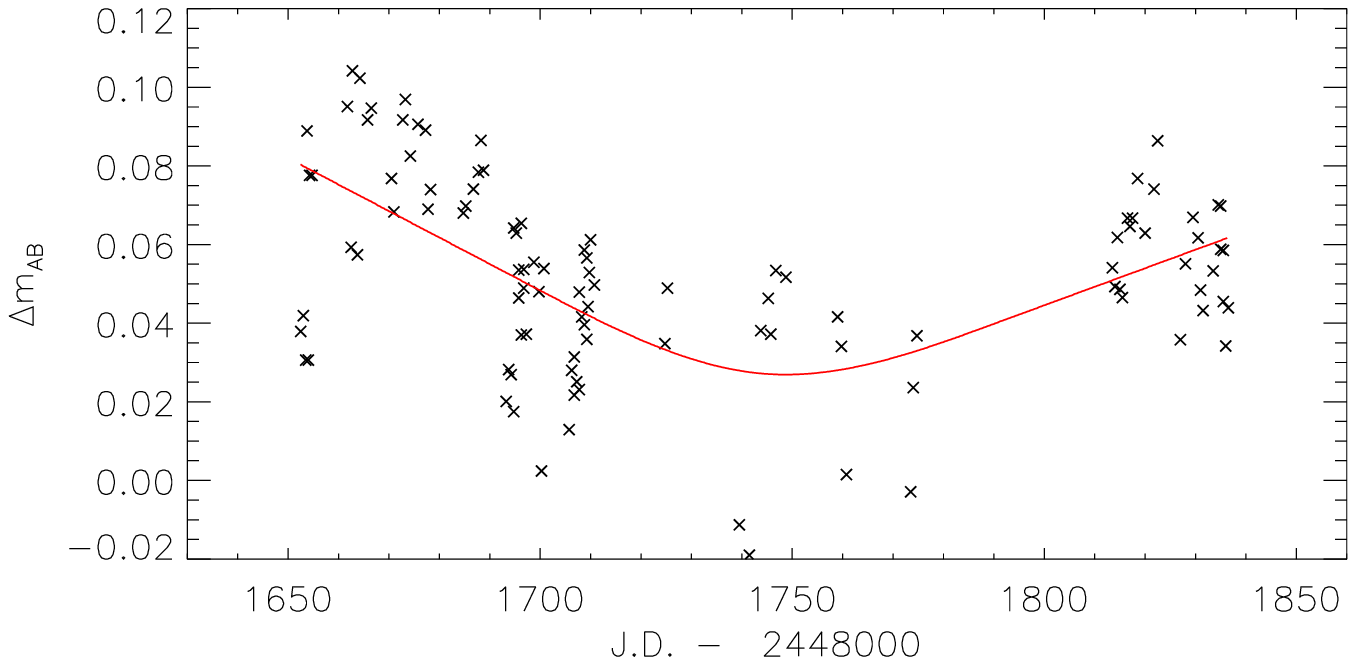}} % regr_fit25_3.ps
\caption{The ML residuals from the three first seasons of
Fig.~\ref{f:AB_residuals}. For two periods we have fitted a cubic
spline (upper panel; 9 nodes, lower panel; 4 nodes).}
\label{f:sub_fits}
\end{figure}

On short time scales (the order of weeks) the residuals are rather
noisy.  Fig.~\ref{f:sub_fits} shows the three first seasons of the ML
residuals in greater detail.
% Figure
For two of the seasons we also include spline approximations. As can
be seen from the middle panel, the data here shows no significant
trends, thus no interpolations are attempted. (Compare the scatter
here to the the mean formal errors in the aperture photometry of the
quasar images, i.e.\ 17~mmag).  It is not trivial to extract
information on the true microlens-induced fluctuations from data such
as these, and the interpolated curves are mostly meant to guide the
eye. However, on the first plot we can discern a steep negative slope
in the residuals around J.D.--2448000~=~900 followed by a positive
slope some 50 days later.  Optimistically, we can explain this as an
``event'' lasting on the order of 70 days, but nothing certain can be
said about its amplitude. It could potentially be a strong event,
because of the steep gradients.  The third period (lower panel)
indicates more clearly a U-shaped feature. The amplitude and time
scale is hard to assess, because we do not have any points in the
``wings''. But is seems that the amplitude is at least 0.05 mag, and
the time scale is on the order of 200 days.

\section{Summary and discussion} \label{S:summary}
We have presented a re-reduction of the CCD image frames critical to
the discussion about time delay and microlensing in the 0957+561
gravitational lens system. Improved computational techniques allow
better subtraction of the effects of the lens galaxy, and correction
for the aperture crosstalk that arises in aperture photometry of the
somewhat overlapping quasar images.

Analysis of the re-reduced photometry for time delay, principally
using several variants of the Dispersion technique, gives consistent
values around 425 days.  The average result from the Dispersion method
and the $\chi^2$ minimization method is 424.9 days, with an estimated
mean error of 1.2 days.  However, we do not claim that the true error
is as small as this. We also note that time delays of roughly 416 to
420 days were \emph{never} seen in this investigation and are thus
less favored by us. This is not in agreement with e.g.\ 
\citet{Kundic97}, \citet{Pelt98} and \citet{CS00}.

Analysis of principally the same image frames with fundamentally
different reduction and time delay estimation techniques had
previously given 404 days (\citealp{Schild97}, Direct Autocorrelation)
and 416.3 days (\citealp{Pelt98}, Dispersion estimation procedure),
but re-analysis by \citet{Oscoz01} of the same brightness record gave
estimates near 422.6 days.  Other smaller data sets for approximately
the same observational epochs gave 417 days (\citealp{Kundic97}, PRH
method, Linear Interpolation) and 425 days (\citealp{Serra99},
$\delta^2$ method).

In all cases but the first, the quoted errors (typically a day or two)
are much smaller than the discrepancies between different data sets or
between estimates for the same brightness record. Critical to the
discussion is the fact illustrated in Fig.~\ref{f:ap_full} that the
FWHM of the dispersion curve has a value of approximately 100 days,
and even the local minima, as seen in Figs.~\ref{f:ap_d1_11_B},
\ref{f:ap_d1_11_p6}, \ref{f:Ap_seasons} and \ref{f:chi_ap}, have a
FWHM of 10--20 days in spite of the daily data sampling and the
available several hundred data points for any test lag
(Fig.~\ref{f:win_ap5}). It is by now evident that something
fundamental limits our ability to estimate time delay to the expected
limits imposed by the data sampling and the observational errors.

The physical origin of this discrepancy has been attributed by
\citet{Colley03} to the fact that the quasar's luminous structure is
time-resolved and microlensed. This combination of microlensing and a
time-resolved source might produce multiple time delays whose pattern
changes from year to year. Some evidence for this may be seen in
Fig.~\ref{f:Ap_seasons}, where the relative importance of persistent
lags near 410, 425, and 430 days seems to have changed during the
observational period. We note, however, that for some subsets these
time lags coincide with minima in the window functions.

This puts a new perspective to the understanding of the role of
microlensing for a quasar source. Previous discussions of microlensing
(see \citealp{Schmidt98}, and references contained therein) have
focused upon the role of small accretion discs crossing the network of
caustics in the magnification diagram produced by MACHOs in the lens
galaxy. If real, microlensing fluctuations on time scales on the order
of 70 days (Sect.~\ref{S:s}) may signal the presence of MACHOs with
masses possibly down to planetary masses.  The small, 5\% amplitude of
this short time scale microlensing signal is consistent with previous
conclusions that the luminous source may be quite large relative to
the Einstein Rings \citep{RS91,RS93,RS97,Refsdal00}.  Because this is
near to the noise level, extremely careful data acquisition and
analysis is called for in determining the time delay and microlensing.

We hope in future papers to extend the time delay and microlensing
analysis, using an even larger data set. A longer observational base
line and maybe more statistical techniques could shed new light on the
time delay issue. We also hope that our new reduction scheme (both
aperture and PSF photometry, see \citealp{Ovaldsen02}) includes some
new features which could be of interest to other researchers.

% *** Acknowledgements and Bibliography *****************************
\begin{acknowledgements}
  We thank J.~Pelt and I.~Burud for kindly providing and explaining
  the computer programs for the time delay estimations. 
\end{acknowledgements}

\bibliographystyle{aa}
\bibliography{biblio}

\end{document}